\documentclass[aps,amssymb,twocolumn,showpacs]{revtex4}

\usepackage{graphicx}

\begin{document}
\title{The Effects of Stacking on the Configurations 
and Elasticity of Single Stranded Nucleic Acids}
\author{A. Buhot and A. Halperin}
\affiliation{UMR5819 (UJF, CNRS, CEA) DRFMC, CEA 
Grenoble, 17 rue des Martyrs, 38054 Grenoble cedex 9, 
France}

\begin{abstract}
Stacking interactions in single stranded nucleic
acids give rise to configurations of an annealed
rod-coil multiblock copolymer. Theoretical analysis
identifies the resulting signatures for long 
homopolynucleotides: A non monotonous dependence 
of size on temperature, corresponding effects on 
cyclization and a plateau in the extension force
law. Explicit numerical results for poly(dA) and 
poly(rU) are presented.
\end{abstract}
\maketitle
Single stranded nucleic acids (ssNA) experience 
stacking interactions~\cite{Cantor}. These favor 
parallel orientation of adjacent aromatic rings 
of the bases giving rise to rigid helical domains. 
Thus far, the possible coupling of stacking and 
the elasticity of ssNA received little attention 
and its existence became recently a subject of 
debate~\cite{Goddard,Aalberts,Ansari}. The issue 
is further complicated because the relevant
thermodynamic and structural parameters reported 
vary widely. In this letter we present a theoretical 
analysis of the configurations and elasticity of
ssNA subject to stacking and identify {\it qualitative} 
effects signalling the coupling of stacking with 
the chains' elasticity. Clear signatures of stacking 
are discernable in {\it long} homopolynucleotides, 
under high salt conditions when loops do not form 
and electrostatic interactions are negligible. 
There are two primary effects. One is the occurrence 
of a minimum in the radius of the chain, $R$, as 
the temperature, $T$, is varied. This leads to 
corresponding effects on the cyclyzation of the 
chains. The second effect is a plateau in the 
extension force law of the ssNA subject to tension 
$f$. The analysis utilizes a model for the helix-coil 
transition in helicogenic polypeptides~\cite{Buhot} 
modified to allow for the weak cooperativity of 
stacking. It focuses on the differences between 
ssNA that stack strongly, polydeoxyadenylate 
(poly(dA)) and polyriboadenylate (poly(rA)), 
and those that exhibit weak or no stacking, 
polyribouridylate (poly(rU)) and polydeoxythymidylate 
(poly(dT)). Such long homonucleotides, comprising 
thousands of nucleotides, can be readily synthesized 
enzymatically~\cite{Bollum}. The results suggest 
that under physiological conditions, these effects 
are important only for poly(dA), poly(rA) and ssNA 
containing large A domains. In the absence of such 
domains, stacking effects become noticeable at $T$ 
lower than $10^{\circ}$C.

The extension force laws of $\lambda$ ssDNA, as 
measured in optical tweezers or AFM experiments, do 
not reveal signatures of stacking~\cite{Smith}. The 
results can be rationalized by considering ssDNA as 
freely-jointed chain characterized by a single Kuhn 
length. This basic picture is augmented to allow for 
loop formation~\cite{Montanari} and for electrostatic 
interactions~\cite{Zhang}. However there is 
no evidence for large $A$ domains in $\lambda$ ssDNA 
and these measurements were carried out in ambient 
$T$ thus precluding, as we shall discuss, significant 
stacking effects. Stacking effects were 
reported~\cite{Goddard,Aalberts,Ansari} in ``molecular 
beacons''~\cite{MB}. These are {\it short} ssDNA 
chains capable of forming stem-loop structures. 
One end carries a fluorophore and the other a 
quencher. Accordingly an open hairpin fluoresces 
and a closed hairpin is quenched. The fluorescence 
intensity and its fluctuations allow to extract 
the fraction of hairpins and the opening and 
closing rate constants. Experiments by the 
Libchaber group revealed differences in cyclization 
behavior of poly(dT) and poly(dA) loops that were
attributed to stacking and its effects on the 
rigidity of the chains~\cite{Goddard,Aalberts}. 
This interpretation was disputed by Ansari et al. 
who ascribed the effects to transient trapping 
of misfolded loops while arguing that both poly(dT) 
and poly(dA) behave as flexible polymers~\cite{Ansari}.
Our analysis does not pertain directly to this 
system since it concerns long chains. However, 
the resulting predictions identify clear signatures 
of stacking when misfolding is not an option.

Stacking involves interactions between nearest 
neighbors and is thus non-cooperative or weakly 
cooperative~\cite{Cantor}. It involves a broad
transition between the stacked, helical state 
obtained at low $T$, and random-coil configurations 
at high $T$. At intermediate $T$, ssNAs comprises
of stacked domains interspaced with ``melted'', 
random-coil ones. The polydispersed domains undergo 
dynamic equilibrium and the overall behavior is 
of annealed rod-coil multiblock copolymer. The 
strength of the stacking interactions vary with
the identity of the bases. It is strongest between 
adenosines (A) and it is weakest among uridines 
(U) and thymines (T). The interactions between
chemically different bases are weak. Thus, stacking 
is most pronounced in poly(dA) and poly(rA) while 
it is weak in poly(rU) and in heteropolynucleotides 
without extended A domains. There is evidence that
stacking does not occur in poly(dT). In every case, 
stacking is insensitive to the concentration of salt. 
We consider ssNA homoplymers comprising $N \gg 1$ 
identical monomers, nucleotides, of which $\theta N$ 
are stacked. The stacked bases form $y N$ helical 
domains. In comparison to the non-stacked bases, the 
excess free energy of each stacked monomer is $\Delta 
f = \Delta h - T \Delta s$. $\Delta h$ reflects the 
enthalpy gain associated with the stacking while 
$\Delta s$ allows for the loss of configurational 
entropy due to the parallel orientation of the 
stacked bases. The reported values of $\Delta h$ 
and $\Delta s$ vary widely. For poly(rA) $\Delta h$ 
ranges between $-3$ to $-10$ kcal/mole while $\Delta s$ 
values span the range $-10$ to $-27$ e.u.. In our 
calculations we will use two sets of values: 
$\Delta h = -13$ kcal/mole and $\Delta s = -40$ e.u. 
as reported for poly(dA) as well as $\Delta h = -2.7$ 
kcal/mole and $\Delta s = -10$ e.u. as reported for 
U stacks~\cite{Cantor}. This choice brackets the range 
of reported parameters and will allow us to set 
tentative boundaries of the experimental conditions 
to explore. The terminal monomers of the domain 
experience stacking interaction with one neighbor 
rather than two. The reduction in their configurational
entropy is possibly weaker. To allow for these two 
effects we assign each terminal monomer with an 
additional free energy $\Delta f_{t}$. The corresponding 
Zimm and Bragg parameters are $s = \exp (-\Delta f/kT)$ 
and $\sigma = \exp (-2\Delta f_{t}/kT)$~\cite{Zimm}. 
The $\theta$ vs $T$ melting curves are broad leading 
to $0.5 \leq \sigma \leq 1$ and suggesting weak
cooperativity. For simplicity we will assume 
perfect non-cooperative behavior with $\sigma = 
1$. For comparison, in helicogenic 
polypeptides the $i$th monomer binds the monomer 
$i+3$ thus giving rise to higher cooperativity
signalled by much smaller $\sigma$ values, of order 
$10^{-2}-10^{-3}$~\cite{Birshtein}.
The distance between the bases in
the stacked form varies between $0.32$~nm and $0.35$~nm 
depending on the measurement technique and $T$. 
In the following we will thus assign a value of 
$b = 0.34$~nm to the projected length of a stacked 
monomer along the axis of the helical domain. 
Because of the non-cooperativity of the stacking the
helical domains are relatively short. While their 
persistence length is not known we will assume that 
it is much longer than the typical domain length
and thus effectively infinite. In contrast we assume 
that the unstacked domains behave as freely jointed 
chains and we neglect excluded volume effects. 
Two parameters thus characterize the coil-like 
domains: the effective length of an unstacked monomer, 
$a$, and the persistence length of the coil, $\lambda$. 
Neither is well established. A common value for $a$ 
is $0.6$~nm~\cite{Smith} while reported values of 
$\lambda$, for stacking-free chains vary between 
$0.75$~nm and $3.5$~nm~\cite{Mills}. We will utilize 
$\lambda = a$ and $\lambda =3.5$~nm. The free energy 
of an unconstrained homonucleotide within this model 
is~\cite{Buhot}: 
\begin{eqnarray}
\frac{F_{0}}{NkT} & = & -\theta \ln s + y \sum_{n}
[p_{s}(n) \ln p_{s}(n) + p_{u}(n) \ln p_{u}(n)]  
\nonumber \\
& - & \sum_{n} [(\mu_{1}^{s} + n \mu_{2}^{s}) 
p_{s}(n) + (\mu_{1}^{u} + n \mu_{2}^{u}) p_{u}(n)]. 
\label{F0} 
\end{eqnarray}
The first term allows for the excess free energy 
of the stacked bases. The next two terms specify 
the mixing entropy arising from the polydispersities
of the stacked and coil-like domains where 
$p_{s(u)}(n)$ is the probability of stacked 
(unstacked) domain comprising of $n$ bases. 
The last four terms impose two constraints by use 
of Lagrange multipliers. $\mu_{1}^{s(u)}$ assures 
the probability normalization while $\mu_{2}^{s(u)}$ 
imposes the average number of monomers, $\theta/y$ 
and $(1-\theta)/y$ respectively, in these 
domains~\cite{Buhot}. The equilibrium conditions, 
$\partial F_{0}/\partial \theta = \partial F_{0}/\partial 
y = 0$, yield $\theta = s/(s+1)$, $y = s/(s+1)^{2} =
\theta (1-\theta)$, $p_{s}(n) = (1-\theta) \theta^{n-1}$
and $p_{u}(n) = \theta (1-\theta)^{n-1}$. The average 
number of bases in a stacked and unstacked domain are 
thus respectively $\langle n \rangle_{s} = \sum_{n} 
n p_{s}(n) = 1/(1-\theta)$ and $\langle n \rangle_{u} 
= \sum_{n} n p_{u}(n) = 1/\theta$.

\begin{figure}[t]
\begin{center}
\includegraphics[width=2.2in]{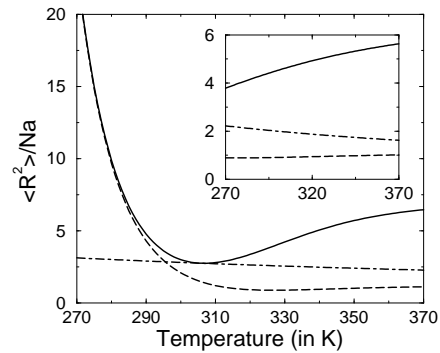}
\caption{Plots of $\langle R^{2} \rangle/N a = 
2 \, l_p$ vs. $T$ for poly(dA) with $l_p$ given 
by Eq.(\ref{lp}) exhibit a minimum at $T_{\min} 
= 307$~K for $\lambda = 3.5$~nm (full line), 
$T_{\min} = 327$~K for $\lambda = 0.6$~nm (dashed 
line) in comparison with $\langle R^{2} \rangle/N a$
vs. $T$ when $l_{p} = \kappa/kT$ (dot-dashed line). 
$\kappa$ was set by equating $l_p$ at $T_{\min}$
for $\lambda = 3.5$~nm. The inset depicts the same 
plots for poly(rU).}
\end{center}
\end{figure}

The simplest characteristic of stacking is the $T$ 
dependence of the mean square radius of the chain, 
$\langle R^{2} \rangle$, as determined by the
end-to-end distance $R$. Assuming that the rod and 
coil segments are freely jointed, the contributions 
of stacked and unstacked monomers to $\langle R^{2}
\rangle $ are independent. The $(1-\theta)N$ 
unstacked monomers in coil domains with a persistence 
length $\lambda$ constitute $N a (1-\theta)/2\lambda$ 
freely jointed segments of length $2\lambda$ contributing 
$(2\lambda )^{2} N a (1-\theta)/2\lambda$ to $\langle 
R^{2} \rangle$. Of the $\theta N$ stacked monomers, 
the ones that form domains incorporating $n$ bases 
contribute $N \theta p_{s}(n)/\langle n \rangle_{s}$ 
freely jointed segments of length $n b$. Altogether, 
\begin{equation}
\langle R^{2} \rangle = N 2 \lambda a 
(1-\theta) + N b^{2} \, \theta \, \langle n^2 
\rangle_s/\langle n \rangle_s
\end{equation}
where $\langle n^2 \rangle_s/\langle n \rangle_s = 
(1+\theta)/(1-\theta)$~\cite{foot1}. The qualitatively 
important feature of $\langle R^{2} \rangle$ is a
minimum (Fig.1) at $\theta_{\min} = 1 - \sqrt{2 
b^{2}/(2 \lambda a + b^{2})}$. This effect disappears 
if we ignore the differences between the size of the
monomer in the two states ($2 \lambda a = b^{2}$). 
It reflects a competition between two contributing 
processes: (i) the number of effective monomers
increases with $T$ because the number and the size 
of the helical domains decrease, and (ii) the size 
of the chain at low $T$ is dominated by a single 
helical domains whose length scales with $N$ rather 
than with $N^{1/2}$. For the values of $a$ and $b$ 
we utilize the minimum is attained at $\theta = 0.769$ 
or $s = 3.32$ for $\lambda = 3.5$~nm and at $\theta = 
0.474$ or $s = 0.9$ for $\lambda = 0.6$~nm. Thus for 
poly(dA) the corresponding $T_{\min}$ values are 
$T_{\min} = 33.6^{\circ}$C and $T_{\min} = 
53.5^{\circ}$C while for poly(rU) $T_{\min} = 
-55^{\circ}$C and $T_{\min} = 2.5^{\circ}$C.
A closely related effect occurs in 
helicogenic polypeptides as they undergo a 
{\it cooperative} helix-coil transition~\cite{Birshtein}.

ssNA chains are occasionally considered as semiflexible, 
worm-like chains. In this case, it is implicitly assumed 
that the properties of the chain along its backbone are 
uniform. In the limit of $N \gg 1$ semiflexible chains 
obey $\langle R^{2} \rangle = 2 l_{p} L$ where $l_{p}$ 
is the persistence length and $L = N a$ is the contour 
length. Within this model $l_{p} = \kappa/kT$ where 
$\kappa = const$ is the bending modulus of the chain. 
If the behavior of the ssDNA is analyzed within this 
framework while the chain obeys the stacking model, 
the $T$ dependence of $l_{p}$ thus extracted is given by 
\begin{equation}
2 l_{p} = 2 \lambda (1-\theta) + (b^2/a) \, \theta \, 
(1+\theta)/(1-\theta). \label{lp}
\end{equation}
At high $T$, when $\theta \rightarrow 0$, $l_{p} \approx 
\lambda$ while for low $T$, when $\theta \rightarrow 1$ 
it diverges as $l_{p} \sim b^{2}/a (1-\theta) \sim b^{2} 
\exp (-\Delta f/kT)/a$. This $T$ dependence is markedly 
different from the $1/T$ behavior predicted by the 
worm-like chain model (Fig.1).

A related signature involves cyclization reactions. 
Cyclization reactions require the two ends of the 
chain to be within a certain capture radius, $r_{c}
\ll \langle R^{2} \rangle ^{1/2}$. The thermodynamics 
of the ring formation are determined by $P(R)dR$, 
the probability for the end-to-end distance of the 
chain to be in the range $R$ to $R+dR$. 
When $N \gg 1$, $P(R)$ of flexible homopolymers, 
behaving as freely jointed chains with constant monomer 
size, assumes a Gaussian form~\cite{Doi}: $P(R) = 
4 \pi R^{2}(2 \pi \langle R^{2}(\theta) \rangle/3)^{-3/2}
\exp (-3R^{2}/2\langle R^{2}(\theta) \rangle)$. 
This result applies to ssDNA in the limit of $\theta 
\rightarrow 0$, when the effect of the stacking is 
negligible. Clearly, it is wrong when $\theta 
\rightarrow 1$ and the configurations are dominated 
by a single, long stacked domain. For $\theta > 0$, 
the Gaussian form is valid provided $N \gg 1$ 
and $y N \gg 1$ i.e., the stacked monomers form a large 
number of domains. The polymer may then be considered as
a freely-jointed chain whose effective monomers are 
rod-coil diblocks of  {\it varying} sizes. The Gaussian 
form is applicable in this regime since the probability 
distribution of lengths of the rod-coil ``monomers'', 
while unspecified, is {\it identical} for all rod-coil 
diblocks~\cite{Chandrasekhar}. The cylization equilibrium 
is ruled by the elastic free energy arising from 
constraining $R$, $F_{el} = - kT \ln P(R) = -kT \ln 
(R^{2}/\langle R^{2}(\theta) \rangle^{3/2}) + 3kT 
R^{2}/2\langle R^{2}(\theta) \rangle + cste$. This 
$F_{el}$ has negligible effect on $\theta$ and $y$ 
because its contribution to the equilibrium conditions 
arises from $R^{2}/\langle R^{2}(\theta) \rangle$. 
For $R \ll N^{1/2}a$ the corresponding terms scale as 
$1/N$ and are thus negligible. To obtain the precise 
cyclization penalty it is important to allow for the 
weighted contributions of all the configurations 
with $R \leq r_{c}$. Since for $r_{c} \ll \langle 
R^{2}(\theta) \rangle^{1/2}$ the exponent in $P(R)$ is 
of order unity, the fraction of cylizable states within 
the freely jointed chain model is $\int_{0}^{r_{c}} P(R)dR 
\sim [ r_{c}/\langle R^{2}(\theta) \rangle^{1/2}]^{3}$. 
For self avoiding chains $\langle R^{2}(\theta) 
\rangle^{1/2}$ is replaced by the $\theta$ dependent 
Flory radius~\cite{DeGennes}. Altogether the cyclization 
entropy for $r_{c} \ll N^{1/2}a$ is $S_{cyc}(\theta) 
= 3k \ln [r_{c}/\langle R^{2}(\theta) \rangle^{1/2}]$ 
and the equilibrium constant for the cyclization reaction 
is specified by $kT \ln K_{cyc} = \Delta H - T 
S_{cyc}(\theta)$ where $\Delta H$ is the binding 
enthalpy of the terminal groups. The activation free
energy for cyclization, $\Delta F_{cyc}^{\ddagger}$ may 
be identified with $-T S_{cyc}(\theta) \sim k T \ln 
\langle R^{2}(\theta) \rangle^{3/2}$~\cite{Aalberts}. 
This $\Delta F_{cyc}^{\ddagger}$ exhibits a minimum at 
$\theta_{\min}$ and thus at the corresponding $T_{\min}$. 
We should emphasize that this analysis is not valid when 
the chain contains a large stacked domain comprising most 
of the monomers. In this case the bending of the chain 
can induce melting of the stacked domain thus introducing 
a coupling of $\theta $ and the cyclization: The elastic 
free energy of fully stacked chain of length $L = l_{p} 
= \kappa/kT$ forming a ring of radius $R = L/2\pi$ is 
$\kappa L/2R^{2} = 2 \pi^{2}kT \approx 20kT$ while the 
reported stacking free energy per base at $25^{\circ}$C 
is in the range of $2$ to $8 kT$. This rough argument 
suggests that the chain can lower its free energy by 
``melting'' a few stacks thus avoiding the bending 
penalty.

When the reaction between the terminal groups is 
diffusion controlled the cyclization rate constant 
assumes the form $k_{cyc} \sim 1/\tau$ where $\tau$
is the longest characteristic time of the chain~\cite{CP}. 
If hydrodynamic interactions are neglected $\tau = 
\tau_{R}$ where $\tau_{R} \approx (\eta_{s} a^{3}/kT) 
N^{2}$ is the Rouse time and $\eta_s$ is the solvent 
viscosity. Allowing for hydrodynamic interaction 
leads to $\tau = \tau_{Z}$ where $\tau_{Z} \approx
\eta_{s} R^{3}/kT$ is the Zimm time. When excluded 
volume interaction are negligible $\tau_{Z} \sim 
N^{3/2}$ while in the opposite case $\tau_{Z} \sim
N^{3\nu}$ where $\nu=0.588$. Experimental studies 
of cyclization of synthetic polymers in non-aqueous 
solutions are consistent with $k_{cyc} \sim N^{-3/2}$. 
Our discussion suggests thus that the cyclization rate
constant of long ssDNA will follow $k_{cyc} \sim 
\langle R^2(\theta) \rangle^{-3/2}$ with a maximum at 
$\theta_{\min}$.

The $T$ dependence of $R$, $K_{cyc}$ and $k_{cyc}$ is 
recovered if we consider the ssDNA as a worm like 
chain with a presistence length $l_p$ given by (\ref{lp}). 
This picture fails qualitatively when considering the 
extension force law. To obtain it, we augment $F_{0}$ 
with the elastic free energy, $F_{el}$, of a 
freely-jointed chain subject to tension $f$~\cite{foot4}. 
A Kuhn length $2 \lambda$ is assigned to the coil 
domains while each of the rigid helical domains orients 
as an independent effective monomer. Allowing for the 
size distribution of the stacked, helical domains we 
obtain~\cite{Buhot} 
\begin{equation}
\frac{F_{el}}{NkT}= -\frac{1-\theta}{\delta} 
{\cal L}_{int}(\delta x) - y \sum_{n}
p_{s}(n) {\cal L}_{int}(\gamma nx)
\end{equation}
where $\delta = 2 \lambda /a$, $\gamma = b/a$,
${\cal L}_{int}(x) = \ln [ \sinh(x)/x]$ and 
$x = fa/kT$. 
Minimization of $F = F_{0} + F_{el}$ in the 
constant $f$ ensemble yields $\theta$, $y$, 
$p_{s(u)}(n)$~\cite{foot5} and the force law 
in the form 
\begin{equation}
\frac{R}{Na} = (1-\theta){\cal L}(\delta x) + 
\theta \gamma \coth (\gamma x) - \frac{\theta 
\gamma sA}{\sinh (\gamma x)} - \frac{y}{x}.  
\label{force}
\end{equation}
At low $T$ the force laws obtained from this model exhibit 
a smoothed plateau associated with the enhancement of stacking 
by the applied force (Fig.2). Initially, the extension lowers 
the entropy of the chain thus favoring the ordered, stacked 
domains. Eventually, further stretching enforces melting of 
the stacks in order to release the stored length ($b=0.34$~nm 
vs $a=0.6$~nm). For our choice of parameters, poly(dA) exhibits 
a pronounced plateau at $T = 25^{\circ}$C while for poly(rU) 
significant deviations from the freely jointed chain appear 
below $T = 10^{\circ }$C. No deviations are expected for 
poly(dT). Modelling the chain as a freely jointed chain with 
a $T$ dependent Kuhn length (\ref{lp}) does not recover this 
plateau because it does not allow for the coupling between 
the stacking and the extension. Accordingly, single molecule 
stretching experiments provide a stringent test allowing to 
distinguish between the conflicting views on the effect of 
stacking on the elasticity of ssDNA.

\begin{figure}[t]
\begin{center}
\includegraphics[width=2.5in]{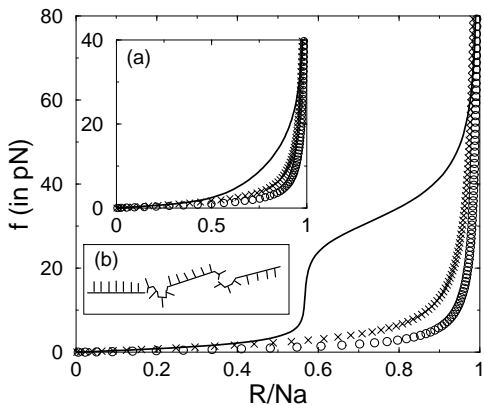}
\caption{Plots of $f$ vs $R/Na$ for poly(dA) at 
$T = 25^{\circ}$C as obtained from the stacking 
model Eq.(\ref{force}) with $\lambda = 3.5$~nm 
(-----), from the freely jointed chain model with 
$l_{p}$ given by Eq.(\ref{lp}) (xxxx) and with 
$l_{p} = 3.5$~nm (oooo). Insets: (a) depicts 
the same plots for poly(rU) at $10^{\circ}$C.
(b) Schematic picture of ssDNA with stacked 
blocks.}
\end{center}
\end{figure}

Straightforward modification of models for the 
helix-coil transition in polypeptides allowed 
us to study the effects of stacking interactions 
on the configurational and elastic properties 
of homonucleotides. In particular, we investigated 
the dependence of $\langle R^{2} \rangle^{1/2}$, 
$K_{cyc}$, $k_{cyc}$ and the force laws, $f$ vs $R$, 
on $T$. The most noticeable effects are extrema 
in the plots of $\langle R^{2} \rangle^{1/2}$ and 
$k_{cyc}$ vs. $T$ as well as the appearance of a 
plateau in the force law. The numerical results are 
based on reported  values of $\Delta h$, $\Delta s$, 
$a$, $b$ and $\lambda$ as obtained from experiments. 
In confronting our predictions with future experiments 
it is important to note that the reported parameters 
vary widely with the measurement technique and the 
experimental conditions. They also depend on the model 
used to analyze the data. 
With this caveat in mind, the qualitative effects we 
discuss provide powerful diagnostics for the coupling 
of stacking interactions with the elastic properties 
of ssNA. These predictions are meaningful because 
spectroscopic evidence indicates significant stacking 
in poly(dA) and poly(rA) at $20^{\circ}$C, irrespective 
of the precise values of $\Delta h$, $\Delta s$, $a$ and 
$\lambda$. The results presented above can also be used 
to test the performance of various sets of $\Delta h$, 
$\Delta s$, $a$, $b$ and $\lambda$ in recovering the 
observable $\langle R^{2} \rangle^{1/2}$ vs $T$ or $f$
vs $R$ plots.


\begin{thebibliography}{99}
\bibitem{foot1} $\langle R^{2} \rangle$ for small 
$\theta$ can be approximated by ignoring 
polydispersity of the helical domains. The ssNA 
comprises thus of $N a (1-\theta)/2 \lambda$ 
unstacked domains of size $2 \lambda$ and 
$N \theta (1-\theta)$ helical domains of length 
$N b/(1-\theta)$. Both the unstacked and helical 
domains are freely jointed leading accordingly to 
$\langle R^{2} \rangle \simeq N 2 \lambda a (1-\theta) 
+ N b^2 \theta/(1-\theta)$.
\bibitem{foot4} For a chain consisting of $N$ 
freely jointed monomers of length $a$, each 
monomer is assigned an energy of $- f a \cos \phi$ 
where $\phi$ is the angle between the axis of the 
segment and the direction of $f$. 
The end-to-end distance satisfies 
$R/Na = {\cal L}(x)$ where ${\cal L}(x) = 
\coth( x)- 1/x$ is the Langevin function and 
$x=fa/kT$. The elastic free energy in the 
constant $f$ ensemble is $-F_{el} = \int_{0}^{f} 
R(f') df'= N kT {\cal L}_{int}(x) = N kT \ln 
[\sinh(x)/x]$.
\bibitem{foot5} $\theta$ and $y$ are solutions of 
$\theta \gamma x (1-sAe^{\gamma x})
(1-sAe^{-\gamma x})=(1-\theta-y)sA(x,\theta,y) \sinh 
(\gamma x)$ and $2\gamma x y =(1-\theta-y) \ln 
[(1-sAe^{-\gamma x})/(1-sAe^{\gamma x})]$ where 
$A(x,\theta,y) = [(1-\theta-y)/(1-\theta)][\delta x/\sinh 
(\delta x)]^{1/\delta}$. The probabilities $p_{s(u)}(n)$ 
are specified by $p_{s}(n) = (1-\theta-y)(sA)^{n} \sinh 
(n \gamma x)/n \gamma x y$ and $p_{u}(n) = 
[(1-\theta-y)/(1-\theta )]^{n} \, y/(1-\theta-y)$.
\end{thebibliography}
\end{document}